\documentclass[twoside]{article}

\usepackage{lipsum} % Package to generate dummy text throughout this template
\usepackage{aas_macros}

\usepackage[sc]{mathpazo} % Use the Palatino font
\usepackage[T1]{fontenc} % Use 8-bit encoding that has 256 glyphs
\usepackage{microtype} % Slightly tweak font spacing for aesthetics

\usepackage[hmarginratio=1:1,top=32mm,columnsep=20pt]{geometry} % Document margins
\usepackage{multicol} % Used for the two-column layout of the document
\usepackage[hang, small,labelfont=bf,up,textfont=it,up]{caption} % Custom captions under/above floats in tables or figures
\usepackage{booktabs} % Horizontal rules in tables
\usepackage{float} % Required for tables and figures in the multi-column environment - they need to be placed in specific locations with the [H] (e.g. \begin{table}[H])
\usepackage{hyperref} % For hyperlinks in the PDF
\usepackage{natbib} %- my add
\usepackage{graphicx}
\usepackage{caption} %subfigures - my add
\usepackage{subcaption} %subfigures - my add
\usepackage{lettrine} % The lettrine is the first enlarged letter at the beginning of the text
\usepackage{paralist} % Used for the compactitem environment which makes bullet points with less space between them
\usepackage{wasysym}
\usepackage{stmaryrd}
\usepackage{abstract} % Allows abstract customization
\usepackage{color}

\usepackage{ulem}

\usepackage{titlesec} % Allows customization of titles
\renewcommand\thesection{\Roman{section}} % Roman numerals for the sections
\renewcommand\thesubsection{\thesection.\arabic{subsection}} % Roman numeralsfor subsections
\titleformat{\section}[block]{\large\scshape\centering}{\thesection.}{1em}{} % Change the look of the section titles
\titleformat{\subsection}[block]{\large}{\thesubsection.}{1em}{} % Change the look of the section titles

\usepackage{fancyhdr} % Headers and footers
\usepackage{color}

\title{\vspace{-15mm}\fontsize{24pt}{10pt}\selectfont\textbf{Lots of Shade on Satellite Constellations}} % Article title

\author{
\large
\textsc{Michael B. Lund$^1$}\\%, Robert J. Siverd$^2$, and Ponder Stibbons$^3$}\\%\thanks{A thank you or further information}\\[2mm] % Your name
\normalsize $^1$Caltech/IPAC-NExScI\\
\normalsize \href{mailto:editor@actaprimaaprilia.com}{editor@actaprimaaprilia.com} % Your email address
\vspace{-5mm}
}
\date{}

\begin{document}

\maketitle % Insert title

\thispagestyle{fancy} % All pages have headers and footers

%----------------------------------------------------------------------------------------
%	ABSTRACT
%----------------------------------------------------------------------------------------

\begin{abstract}

The high frequency of satellite launches, particularly over the last few years, has been a subject of significant concern, particularly relating to the future of observational astronomy, the stability of low Earth orbits, and environmental impacts. We call attention to the insufficiently-addressed silver lining of this looming satellite cloud. If the high rates of satellites continue as we model, we can expect the solar flux received by the Earth to significantly decrease in the relatively near future. We address how this decrease in flux could provide a solution for another major problem, anthropogenic climate change. This would allow us to solve one problem with another problem as early as late March 2031.
\end{abstract}

%----------------------------------------------------------------------------------------
%	ARTICLE CONTENTS
%----------------------------------------------------------------------------------------

\begin{multicols}{2} % Two-column layout throughout the main article text

\section{Introduction}
\lettrine[nindent=0em,lines=3]{"S}
ince the beginning of time, man has yearned to destroy the sun. I will do the next best thing...block it out!" - Anonymous Philanthropist\footnote{\citep{Oakley1995}}

Two concerning human-caused trends have been taking place, having received significant attention from scientists broadly and, in many cases, astronomers and physicists in particular. The first is that the massive increase in greenhouse gas production has driven climate change at what is becoming a concerning rate. The second is that the veritable explosion of satellites in low Earth orbits has very real consequences on both the environment, the future of astronomy, and humanity's connection to the stars. These could be viewed as two very serious problems, or we could take the approach that "every problem is just a solution waiting to be found" \citep{Ashby2010}.

Consequently, while there has been some discussion about how sunshades could be used as a component in active geoengineering to lower planetary temperatures \citep{Wolverton1994}, and while we could do the work to study this carefully and cautiously before engaging in any widespread efforts, the rate of active satellites could free us from that burden. Instead, we may be able to rely on a near future where satellites are plentiful enough to have a meaningful impact on global warming.

In Section~\ref{Background}, we recap both the understanding of the sun's impact on the Earth's climate and some of the discussion on the impacts of satellites, particularly those in low Earth orbits. In Section~\ref{Methods} we briefly discuss our source for the number of active satellites and outline our approach using this data to assess sky coverage as a function of time. Finally, we summarize this paper in Section~\ref{Summary}.

%------------------------------------------------
\section{Background} \label{Background}

\subsection{Sun on Earth}\label{Sun}
It is not a groundbreaking result to suggest that the sun has an impact on the Earth's climate, something carried out predominantly through radiation \citep{Planck1914}. Indeed, even the nuances of this were being explored incredibly early on, such as the early work that suggested that the Earth's atmosphere was keeping temperatures higher than it would otherwise be in the early 19th century \citep{Fourier1824}. By the end of that century, the first paper to quantify how carbon dioxide contributed as a greenhouse gas and to discuss how variations in greenhouse gases would impact climate by the end of that century \citep{Arrhenius1896}. Substantially more recently, the most recent report from the Intergovernmental Panel on Climate Change continues to point toward changes in atmospheric composition caused by humans as the dominant driver of climate change today \citep{Calvin2023}.

While the anthropogenic release of greenhouse gases currently gets the bulk of discussion on what drives major climate changes (for obvious reasons), extraordinary circumstances can lead to other phenomena causing temperature changes on Earth. One of the most clear examples is the observed drops in temperature that occur due to solar eclipses, when the moon eclipses the sun from the perspective of the Earth \citep{Aplin2016}. An arguably even more significant change in flux occurs when the Earth eclipses the sun from the perspective of the earth, though we have not found this expressly explored in the literature, despite checking nightly.

Smaller phenomena than the moon can also impact global temperatures by stopping some portion of light before it can reach the Earth's surface. Of course, this process begins at the source. Although the IAU has defined specific values for things such as solar irradiance and effective temperature in recent years \citep{Prsa2016}, a climate data record for solar irradiance that captures its variability over hundreds of years has also been developed \citep{Coddington2016}. It remains a topic of discussion just how influential this variation has been for terrestrial temperatures in the last millennium, but variations in solar irradiance appear to have had a smaller impact on global temperature than changes in greenhouse gases and volcanic activity over the same timeframe \citep{Hegerl2018, Schurer2014}.

Away from the source and closer to home, the aforementioned volcanic events have also been able to impact global warming by preventing more of the light that reaches the Earth from entering the atmosphere and triggering, though according to \citep{Bradley1988} the magnitude of cooling caused by individual eruptions generally reduces down to the level of the noise on a time scale of only 1-2 years. In some level of contrast, \citep{Briffa1998} found that clusters of volcanic activity could cause cooling on multidecadal levels "as occurred in the midfifteenth century, the 1640s, 1660s and 70s, 1690s and the 1810s" and that the clustering of eruptions in the middle of the 17th century may be responsible for extended hemispheric cooling.

An example of these competing causes is found in the study of the Little Ice Age, a period of cooling from roughly the 14th to 19th centuries, which has been discussed both in the context of being driven by volcanic activity with no change in solar irradiance needed \citep{Miller2012} and within the framework of solar variability as a cause for the period of cooling \citep{Lean1999}.

Addressing our more modern concerns on how greenhouse gases have been responsible for rising global temperatures, \citep{Govindasamy2000} found that the temperature increases caused by a doubling in carbon dioxide levels would be largely counteracted by a solar dimming of just 1.8\%.

\subsection{Satellite Impacts}\label{Satellites}
There has been a significant amount of discussion about the impact of satellites on the Earth. This shouldn't be confused with the impact of satellites with one another, though that is also a rich topic of study, starting from \citet{Kessler1978} proposing what is now known as the Kessler Effect, where the density of satellites is sufficiently high that one collision between two orbiting satellites causes a cascading cloud of destruction as a chain of uncontrolled collisions takes place.
A more recent assessment of what properties of satellites are more likely to contribute to the risk of chain reaction collisions taking place was carried out by \citet{Ballard2025}, identifying apogee and orbital period as two of the larger risk factors. There has also been a proposal for a new metric, called the CRASH Clock, which is how long it would take for a catastrophic collision to occur if avoidance maneuvers could not be taken or there was a loss of situational data to assess that there is a risk \citep{Thiele2025}. According to \citet{Thiele2025}, the CRASH Clock has decreased from about 120 days in 2018 to just under 3 days in 2025; this suggests that megaconstellations of satellites are now very vulnerable to temporary disruptions that may trigger massive long-term consequences. The United States Federal Aviation Authority has even found that the risk to planes of being struck by space debris will increase by more than an order of magnitude from 2021 to 2035 \citep{FAA2023}. As concerning as this all may sound, in this paper we are more concerned with the impact of satellites from the perspective of the Earth's surface.

Specifically in the context of astronomy, there has been significant concern on the impact that satellites have on observations of the night sky \citep{NatAstro2026}. Individual satellites have been concerning enough, such as prototype satellite BlueWalker 3's $64 m^{2}$ antenna making it one of the brightest objects in the sky, reaching a magnitude of +0.4 \citep{Nandakumar2023}.

Serious efforts to call for safeguarding the night sky from increasing numbers of satellites can go back at least as far as \citep{Gallozzi2020}, which argued for a stop to constellations of satellites that were going to jeopardize astronomy as a whole, including the risk that this may pose to planetary defense. More targeted assessments have included looking at the impact that satellites will have on observations with both the Hubble Space Telescope \citep{Kruk2023} and the Vera C. Rubin Observatory's Legacy Survey of Space and Time (LSST) \citep{Kandula2025}, while a broader assessment of the contamination that will impact current and new-generation space-based telescopes as a class is available in \citep{Borlaff2025}.

The impact of these satellites (and the fundamental processes involved in launching this many satellites) are not limited to contamination observations. For example, the increasing number of rocket launches is becoming ever more damaging to the ozone layer \citep{Dallas2020, Revell2025}. This would represent a significant step backwards given the important success of global buy-in to the Montreal Protocol after damage to the ozone layer was discovered in the 1980s \citep{Solomon1986, Wang2025}. There are additional questions that remain regarding the anthropogenic injection of aerosols into the atmosphere as satellites and rocket bodies burning up upon reentry will soon begin to outpace naturally-occuring meteroids as the dominant source of metals in the upper atmosphere \citep{Schulz2021}.

%------------------------------------------------
\section{Data and Methods} \label{Methods}
\subsection{Data}
For the number of active satellites orbiting the Earth as a function of time, we have used the data that have been collected on \emph{Jonathan's Space Pages
}\footnote{https://planet4589.org/space/stats/active.html}. At the time of writing, this dataset covers the number of active satellites from December 31, 1956 to March 18, 2026.

\begin{figure*}[!htb]
  \begin{center}
   \includegraphics[width=0.95\textwidth]{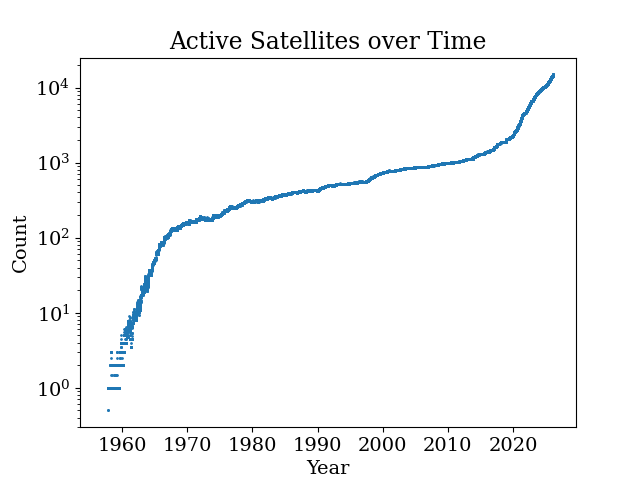}
  \end{center}
  \caption{Active satellites as a function of time. The multiple bends when plotted logarithmically indicate a single exponential function can't fit this data.}
  \label{fig:data_log}
\end{figure*}

For computational purposes, we downsample our dataset from daily data points to just the number of active satellites for January 1 of each year. We show this data in Figure~\ref{fig:data_log}, where the data cannot be characterized by a single period of linear or exponential growth.

\subsection{Analysis}
While we can still try an exponential fit to the data, we also can see that an exponential does not adequately address how the number of active satellites is relatively low until it begins to increase a prodigious rate within the last decade or so. This reflects how this is actually superexponential growth, which is also shown by how this looks on a logarithmic scale such as in Figure~\ref{fig:data_log}.

Instead, we can better fit this by treating as J-curve growth. Such an approach to capture rapid explosions in a measured quantity and predict future values has been carried out in the past. \citep{Martino1972} discusses how, in 1952-1953, the US Air Force used this approach to look at maximum speeds in a way that successfully anticipated mankind reaching the speeds necessary for a satellite in orbit in 1957 (Sputnik \citep{Science1957}) and a probe at escape velocity in 1959 (Luna 1). Another example of use of a J-curve was by \citet{vonFoerster1960}, in a work titled "Doomsday: Friday, 13 November, A.D. 2026". Recent observations suggest that von Foerster may have been quite prescient in that prediction.

For the sake of a more familiar comparison, we begin by looking at a standard exponential function of the form in Equation~\ref{eq:exp}, where t is the year and our resulting value f(t) is the number of active satellites in orbit around the Earth.
\begin{equation} \label{eq:exp}
f(t) = A(B)^{t-C}
\end{equation}

To address the extreme growth observed, the equation that we are more interested in using to capture superexponential growth is Equation~\ref{eq:superexp}, where t is the year and again our resulting value f(t) is the number of active satellites in orbit around the Earth.
\begin{equation} \label{eq:superexp}
f(t) = A/(B-t)^{C}
\end{equation}

\begin{figure*}[!htb]
  \begin{center}
   \includegraphics[width=0.95\textwidth]{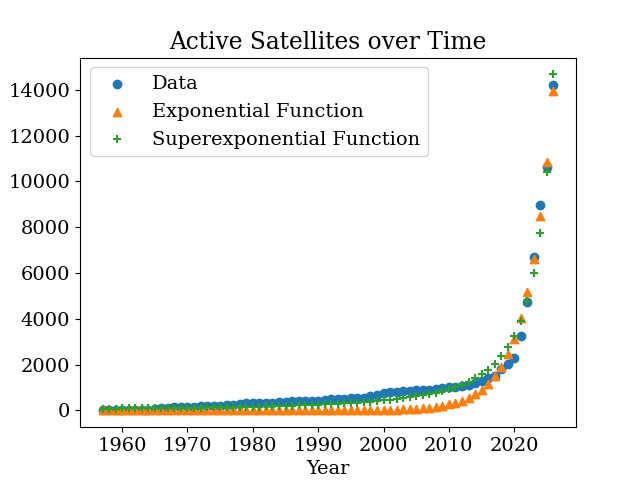}
  \end{center}
  \caption{Active satellites as a function of time. Data is represented by blue circles, while the best-fit exponential and superexponential functions are represented by orange triangles and green crosses, respectively.}
  \label{fig:func_data}
\end{figure*}

Our best-fit exponential function is for the values A = 4.29198, B = 1.28213, and C = 1993.46 with an $R^{2}$ value of 0.964. However, the superexponential function's $R^{2}$ value is 0.984, with values of A = 384177, B = 2031.24, and C = 1.97231. This shows that the superexponential function better characterizes the available data, This can also be seen in Figure~\ref{fig:func_data}, largely because the exponential function fails to model the rise in the number of active satellites that takes place from the 1980s through to the mid-2010s.

The quantity that we are truly interested in, however, is not the number of active satellites but rather the amount of the sky that those satellites block out. As over two thirds of active satellites are specifically Starlink\footnote{Jonathan's Space Statistics - see Footnote 2}, we use nominal values for Starlink satellites as our approximations for all active satellites, specifically a semi-major axis of 6750 km \citep{Lang2025} and a cross-sectional area of 5 square meters \citep{Ahmad2025}. From those two values and the number of active satellites, we can calculate the fraction of the surface of a sphere with radius of 6750 km the total number of active satellites would cover. Rearranging that equation, we can instead find the time at which we reach a desired fractional coverage, f. In Equation~\ref{eq:solve_t}, we can solve for t to find when we will reach a given fractional coverage for specified values of the orbital radius, r, cross-sectional area of a satellite, a, and the values of A, B, and C from our fit for the superexponential function in Function~\ref{eq:superexp}. Our fractional coverage threshold is 0.018, as discussed earlier in \citet{Govindasamy2000}.

\begin{equation} \label{eq:solve_t}
t = B - \left(\frac{Aa}{4f\pi r^{2}} \right)^{1/C}
\end{equation}

When all values are plugged into this equation, we find that the amount of sky covered by active satellites will reach our 1.8\% coverage threshold in 2031. Specifically,we find a value of 2031.2355, which corresponds to shortly before midnight on March 27th, 2031.

We then have until 2031 to figure out a solution to the next challenge, which is that at this rate of skyrocketing satellite output, we would reach this threshold only a few hours before reaching full Earth containment with satellites covering 100\% of the available sky, setting our new Doomsday to Friday, 28 March, A.D. 2031.

%------------------------------------------------

\section{Summary}\label{Summary}
In this work, we have suggested that one major problem, antropogenic climate change, maybe be readily solved by another of our major problems, the rapid proliferation of satellites resulting in increasingly crowded low Earth orbits. Studies on the Sun's impact on the Earth and on climate change in general have suggested that a 1.8\% dimming in solar flux could effectively counteract a significant component of climate change's warming due to the recent increase in carbon dioxide in the atmosphere.

Concurrently, the number of active satellites is growing at such a rate that humanity's history of space exploration cannot be sufficiently modeled by exponential growth. Using a more suitable superexponential growth model, we show that there will be enough satellites in orbit, blocking out enough of the sky, that the amount of sunlight reaching the Earth could drop significantly enough in the next five years to counteract global warming trends.

\section{Acknowledgements}
%This research made use of Astropy,\footnote{http://www.astropy.org} a community-developed core Python package for Astronomy \citep{astropy:2013, astropy:2018}.

This research has made use of NASA’s Astrophysics Data System.

This research has made use of the Science Explorer, funded by NASA under Cooperative Agreement 80NSSC21M00561.

This research has made use of data collected for \emph{Jonathan's Space Pages} by Jonathan McDowell.

Software: matplotlib \citep{Hunter2007, Caswell2023}, numpy \citep{Harris2020}, pandas \citep{Mckinney2010, Reback2020}, scipy \citep{Virtanen2020}

%----------------------------------------------------------------------------------------
%	REFERENCE LIST
%----------------------------------------------------------------------------------------

\bibliographystyle{apalike}
\bibliography{main}

%----------------------------------------------------------------------------------------

\end{multicols}

\end{document}